\documentclass[12pt]{article}
\usepackage{graphicx}
\usepackage{subfigure}
\usepackage{amsfonts}
\usepackage{amssymb,amsmath}

\newcommand{\nn}{\nonumber \\}
\newcommand{\bea}{\begin{eqnarray}}
\newcommand{\eea}{\end{eqnarray}}

\makeatletter
\@addtoreset{equation}{section}

\makeatother

\usepackage{multicol}
\usepackage[usenames,dvipsnames]{xcolor}
\definecolor{rosso}{cmyk}{0,1,1,0.3}
\definecolor{verde}{cmyk}{0.8,0,0.6,0.25}
\definecolor{bluc}{cmyk}{1,0.4,0,0.1}
\definecolor{blucc}{cmyk}{0.8,0.3,0,0}

\def\be{\begin{equation}}
\def\ee{\end{equation}}

\def\({\left(}
\def\){\right)}
\def\1{^{(1)}}
\def\2{^{(2)}}

\def\<{\langle}
\def\>{\rangle}

\begin{document}

\begin{titlepage}

\begin{flushright}
UT-13-03\\
IPMU 13-0024
\end{flushright}

\vskip 3cm

\begin{center}

{\Large \bf 
Axino dark matter with R-parity violation and\\
130 GeV gamma-ray line
}

\vskip .5in

{
Motoi Endo$^{(a,b)}$,
Koichi Hamaguchi$^{(a,b)}$,
Seng Pei Liew$^{(a)}$,\\
Kyohei Mukaida$^{(a)}$
and
Kazunori Nakayama$^{(a,b)}$
}

\vskip .3in

{\em
$^a$Department of Physics, University of Tokyo, Bunkyo-ku, Tokyo 113-0033, Japan \vspace{0.2cm}\\
$^b$Kavli Institute for the Physics and Mathematics of the Universe,
University of Tokyo, Kashiwa 277-8583, Japan \\
}

\end{center}

\vskip .5in

\begin{abstract}

We show that decaying axino dark matter with R-parity violation can explain the 
observed excess of the 130GeV gamma-ray line from the Galactic center in the Fermi data.
The branching fraction of the axino decay into monochromatic photons can be $O(1)$, and  
constraints from continuum gamma-rays and the anti-proton flux are ameliorated.
The Peccei-Quinn scale of  $O(10^{13}-10^{14})$\,GeV and the R-parity violation parameter of $O(10^{-12}-10^{-11})$
are cosmologically favored.

\vskip.5in
\noindent

\end{abstract}

\end{titlepage}

\setcounter{page}{1}

\section{Introduction} 
\label{sec:introduction}

Recently, there is increasing evidence of the excess of the 130\,GeV gamma-ray line from 
the Galactic Center (GC) in the four-year Fermi data~\cite{Bringmann:2012vr, Weniger:2012tx,Tempel:2012ey,Boyarsky:2012ca,Su:2012ft,Aharonian:2012cs,Hektor:2012kc,Hektor:2012ev,Finkbeiner:2012ez}.
This may be interpreted as a signal of the dark matter (DM), which annihilates or decays around the GC.
An obstacle to construct a model of the annihilating/decaying DM which explains the observed gamma-ray line excess is 
that the branching ratio of the monochromatic photon production must be fairly large. It should be larger than around 0.01~\cite{Buchmuller:2012rc,Cohen:2012me,Cholis:2012fb}. 
Otherwise, continuum gamma-rays would hide the line gamma, and anti-protons may be overproduced.
For instance, if the DM annihilation into photons takes place through loops of the standard model (SM) particles, it is difficult to realize such a large branching ratio~\cite{Asano:2012zv}.

In this letter, we propose a model of the decaying DM which naturally explains the gamma-ray line excess without producing too much continuum gammas and anti-protons.
A supersymmetric (SUSY) axion model~\cite{Kawasaki:2013ae} is considered 
to solve the strong CP problem in the framework of the minimal SUSY SM (MSSM).
The axino, which is a fermionic superpartner of the axion, is a suitable candidate of the DM,
if it is the lightest SUSY particle (LSP).
By introducing small R-parity violations, the axino decays into a photon plus a neutrino,
and the Fermi gamma-ray line excess can be explained. 
It is stressed that the branching fraction of the axino decay into monochromatic photons typically becomes $O(10)\%$, and 
the constraints from the overproductions of the continuum gamma-ray and the antiproton are satisfied.
This is in contrast to the decaying gravitino DM scenario,
where the branching fraction of the monochromatic photon production is suppressed~\cite{Buchmuller:2012rc}.
Moreover, the present scenario is cosmologically favored, because the lightest SUSY particle of the MSSM (MSSM-LSP), e.g., the lightest neutralino, decays by the R-parity violating effects
before the big-bang nucleosynthesis (BBN) begins.
This avoids the cosmological problem associated with a late decay of the MSSM-LSP when the gravitino is lighter than the MSSM-LSP.\footnote{
	Light axino DM with R-parity violations was considered in Refs.~\cite{Kim:2001sh,Hooper:2004qf,Chun:2006ss}.
	Ref.~\cite{Hasenkamp:2011xh} considered the gravitino LSP with the axino heavier than MSSM-LSP.
}
On the other hand, the morphology of the gamma-ray line signature from the GC seems to favor the annihilating DM scenario rather than
that of the decaying DM~\cite{Buchmuller:2012rc}.
Although relatively large gamma-ray signals are expected from the Galactic halo in the decaying DM scenario, 
no such excesses have been observed. 
However, since there are potentially large uncertainties in the gamma-ray data and the DM density profile around the GC, 
it is premature to specify the DM model by the morphology~\cite{Buchmuller:2012rc,Park:2012xq}.

In the next section, the axino DM model will be introduced, and properties of the model will be explained, particularly paying attention to the R-parity violating effects. We consider the KSVZ axion models~\cite{Kim:1979if}. It will be shown that the model can explain the gamma-ray line excess. In addition, several cosmological aspects will be discussed. The last section will be devoted to the conclusion and discussion.

\section{Axino dark matter with R-parity violation} 
\label{sec:axino}

\subsection{Axino decay rate with R-parity violation} \label{sec:R}

Let us first introduce R-parity violations. 
In this letter, we consider a bilinear type of the 
R-parity
violation~\cite{Barbier:2004ez}, 
which is characterized by the superpotential,
\begin{equation}
	W = \mu_i L_i H_u,  \label{W_bi}
\end{equation}
where $L_i$ and $H_u$ are chiral superfields of the lepton doublet and the up-type Higgs doublet, respectively.
The index $i=1,2,3$ denotes the generation, and $\mu_i$ is a parameter with a mass dimension.
Here and hereafter, summation over $i$ is implicitly promised.
By redefining $L_i$ and the down-type Higgs superfield $H_d$ as
$L_i' = L_i - \epsilon_i H_d$ and $H_d' = H_d +\epsilon_i L_i$ with $\epsilon_i \equiv \mu_i/\mu$,
where $\mu$ is the higgsino mass parameter appearing in the superpotential as $W = \mu H_u H_d$,
the R-parity violating superpotential (\ref{W_bi}) is eliminated.
Hereafter, for notational simplicity, the primes on the redefined fields are omitted.
After the redefinition, the SUSY breaking potential becomes
\begin{equation}
	-\mathcal L_{\rm RPV} = B_i \tilde L_i H_u + m_{L_i H_d}^2 \tilde L_i H_d^* +{\rm h.c.},  \label{LRPV}
\end{equation}
where $\tilde L_i$ is a scalar component of the superfield $L_i$. The coefficients are $B_i \simeq - B \epsilon_i$ and 
$m_{L_i H_d}^2 \simeq (m_{\tilde L_i}^2 - m_{H_d}^2 )\epsilon_i$, where $B$, $m_{\tilde L_i}^2$
and $m_{H_d}^2$ represent soft SUSY breaking parameters in the MSSM, 
$-\mathcal L_{\rm soft} = (BH_uH_d+{\rm h.c.}) + m_{H_d}^2 |H_d|^2 + m_{H_u}^2 |H_u|^2 
+ m_{\tilde{L_i}}^2 |\tilde{L_i}|^2 + \cdots$.
Due to the R-parity violating scalar potential (\ref{LRPV}), sneutrinos obtain non-zero vacuum expectation values (VEVs) as
\begin{equation}
	\langle \tilde \nu_i\rangle = -\frac{m_{L_i H_d}^2\cos\beta + B_i\sin\beta}{m_{\tilde\nu_i}^2}v,
	\label{nuVEV}
\end{equation}
where $\tan\beta \equiv v_u/v_d$ is a ratio of the VEVs of the up- and down-type Higgs fields,
$v \equiv \sqrt{v_u^2+v_d^2}\simeq 174$\,GeV, and $m_{\tilde\nu_i}^2$ is a sneutrino mass.

Before proceeding to discuss phenomenological aspects, several comments are in order. It is possible to introduce the bilinear R-parity violating soft terms, $\tilde L_i H_u$ and $\tilde L_i H_d^*$ in addition to \eqref{W_bi}, before the field redefinition. The coefficients in \eqref{LRPV} 
then have additional contributions, but the following analysis will not be affected
as far as the R-parity violation is parametrized by the the sneutrino VEV \eqref{nuVEV}.
Next, trilinear R-parity violating terms, $LLE$ and $LQD$, are also generated by the field redefinition. They are subdominant and will be ignored in the following study, because the terms are multiplied by the Yukawa couplings.

The sneutrino VEVs (\ref{nuVEV}) induce mixings between the SM leptons and the gauginos. 
The SM neutrinos mix with the bino and the neutral wino,
and the SM charged leptons mix with the charged winos.
Hence, the R-parity violating parameters are constrained. 
The neutrinos obtain masses of $m_\nu \sim g^2 \langle \tilde \nu\rangle ^2 / m_{\tilde {B}(\tilde{W})}$, 
where $m_{\tilde {B} (\tilde {W})}$ is a bino (wino) mass~\cite{Romao:1999up,Takayama:1999pc,Hirsch:2004he}.
For gaugino masses of $O(100)$\,GeV, $\kappa_i \equiv \langle \tilde \nu_i\rangle/v \lesssim 10^{-7}$ is imposed 
to satisfy the experimental bound on the neutrino masses.
Also, the cosmological $B-L$ asymmetry is preserved for $\kappa_i \lesssim 10^{-7}$~\cite{Campbell:1990fa,Fischler:1990gn,Dreiner:1992vm,Endo:2009cv}. 
Other constraints are known to be weaker (see e.g., Ref.~\cite{Barbier:2004ez}). 
As we will see, the size of the R-parity violation favored by the Fermi gamma-ray line excess is much smaller as
$\kappa_i\sim 10^{-11}$.
\footnote{
See e.g., Refs.~\cite{Buchmuller:2007ui,Endo:2009by,Bobrovskyi:2010ps} for models to explain such a tiny R-parity violation parameter.
}

The R-parity violation destabilizes the LSP. 
In this letter, we consider the axino LSP scenario in the KSVZ axion models~\cite{Kim:1979if}. 
The relevant interaction terms of the axino are
\begin{equation}
	 \mathcal{L}_{\tilde{a}\lambda A}= i \frac{\alpha_Y C_Y}{16 \pi f_a}\bar{\tilde{a}}
	 \gamma_5[\gamma^{\mu},\gamma^{\nu}]\tilde{B}B_{\mu \nu} + i \frac{\alpha_W C_W}{16 \pi f_a}
	 \bar{\tilde{a}}\gamma_5[\gamma^{\mu},\gamma^{\nu}]\tilde{W}^aW^a_{\mu \nu}
	\label{aYY}
\end{equation}
where $C_Y$ and $C_W$ are model-dependent coupling constants of order unity,
$\alpha_Y (\equiv g_Y^2/4\pi)$ is the fine structure constant of U(1)$_Y$, 
$\alpha_W (\equiv g_2^2/4\pi)$ is that of SU(2),
$f_a$ is the PQ scale,
$\tilde a$ denotes the axino, $\tilde B (\tilde W^a)$ is the bino (wino),
and $B_{\mu\nu} (W_{\mu\nu})$ is the field strength of the U(1)$_Y$ (SU(2)) gauge boson.
The axino LSP is stable as long as the R-parity is conserved, whereas it decays via the operators \eqref{aYY}
with the gaugino mixings with the SM leptons, once the R-parity violation is turned on. 

First, let us consider the case of $C_W = 0$ (see Sec.~\ref{sec:model} for an explicit realization).
The first term in (\ref{aYY}) provides interactions of $\tilde a - \tilde B - \gamma$ and $\tilde a - \tilde B - Z$.
The R-parity violation opens a decay of the axino through the $\tilde B - \nu_i$ mixing as 
$\tilde a \to \nu_i \gamma$ and $\tilde a \to \nu_i Z$.
In the limit of $|m_{\tilde B} - \mu| \gg m_Z$, the axino decay rate becomes
\begin{equation}
	\Gamma(\tilde a \to \gamma \nu_i) \simeq \frac{C_Y^2\alpha_Y^2}{128 \pi^3}\frac{m_{\tilde a}^3}{f_a^2}
	\left(\frac{g_Y^2 \langle\tilde \nu_i\rangle^2}{2m_{\tilde B}^2}\cos^2\theta_W \right),
	\label{gamma}
\end{equation}
where $m_{\tilde a}$ is the axino mass, and $\theta_W$ is the weak mixing angle.
Here and hereafter, $\Gamma(\tilde a \to \gamma\nu_i / Z\nu_i)$ denotes a sum of the partial decay rates into $\nu_i$ and $\bar{\nu_i}$.
The factor $(g_Y\langle\tilde \nu_i\rangle/\sqrt{2} )/m_{\tilde B}$ in the parenthesis comes from the bino-neutrino mixing, and $\cos\theta_W$ from the U(1)$_Y$ gauge boson-photon mixing.
Similarly, we obtain
\begin{equation}
	\Gamma(\tilde a \to Z \nu_i) \simeq \frac{C_Y^2\alpha_Y^2}{128\pi^3}\frac{m_{\tilde a}^3}{f_a^2}
	\left(\frac{g_Y^2 \langle\tilde \nu_i\rangle^2}{2 m_{\tilde B}^2}\sin^2\theta_W \right)
	\left(1- \frac{m_Z^2}{m_{\tilde a}^2} \right)
	\left( 1-\frac{m_Z^2}{2m_{\tilde a}^2} - \frac{m_Z^4}{2m_{\tilde a}^4} \right),
\end{equation}
where $m_Z$ is the mass of the $Z$ boson.
For $m_{\tilde a} \gg m_Z$, the branching fractions are given by
${\rm Br} (\tilde a \rightarrow \gamma \nu):{\rm Br} (\tilde a \rightarrow Z\nu) \simeq 
\cos^2 \theta_W : \sin^2 \theta_W$.
From the above results, the axino lifetime is estimated as
\begin{equation}
	\tau_{\tilde a}\simeq 8 \times 10^{26}\,{\rm sec}~ C_Y^{-2} 
	\left( \frac{m_{\tilde a}}{260\,{\rm GeV}} \right)^{-3}
	\left( \frac{f_{a}}{10^{13}\,{\rm GeV}} \right)^{2}
	\left( \frac{m_{\tilde B}}{1\,{\rm TeV}} \right)^{2}
	\left( \frac{\kappa}{10^{-11}} \right)^{-2},
\end{equation}
where the R-parity violating parameter is defined as $\kappa \equiv \sqrt{\sum_i \langle\tilde \nu_i\rangle^2} / v$.

The two-body decay of the axino into a photon contributes to the monochromatic gamma signal of the Fermi observation. If the axino mass is around 260\,GeV, the photon produced by $\tilde a \to \gamma\nu$ has an energy of about 130\,GeV.
According to Ref.~\cite{Buchmuller:2012rc}, the observed excess of the gamma-ray line is accounted for by a decaying DM, 
when its lifetime $\tau_{\rm DM}$ and the branching ratio  ${\rm Br}({\rm DM}\to \gamma\nu)$ are in the range of $\tau_{\rm DM}/{\rm Br}({\rm DM}\to \gamma\nu) = (1-3)\times 10^{28}$\,sec and ${\rm Br}({\rm DM}\to \gamma\nu)\simeq 0.01 - 1$. The astrophysical constraints from the diffuse gamma-rays~\cite{Ackermann:2012rg} and neutrinos~\cite{Covi:2009xn} are also satisfied for such a parameter region.
In the present model, the branching ratio is around 0.8 for $C_W = 0$, while the hadronic branch from $\tilde a\to Z\nu_i$ is sufficiently small.
Thus, the lifetime and the branching fraction which are required to explain the gamma-ray line excess from the GC are realized by the axino DM.

Next, let us focus on the case of $C_W\neq 0$ and neglect the contribution from the first term in (\ref{aYY}), i.e., $C_Y = 0$.
The second term in (\ref{aYY}) provides interactions of $\tilde a - \tilde W^0 - \gamma$, $\tilde a - \tilde W^0 - Z$ and $\tilde a - \tilde W^{\pm} - W^{\mp}$.
The decays, $\tilde a \to\nu_i \gamma , \nu_i Z$ and $l_i^{\pm} W^{\mp}$, proceed by these interactions with the mixings of $\tilde W^0 - \nu_i$ and $\tilde W^\pm - l_i^{\pm}$.
In the limit of $|m_{\tilde W} - \mu|\gg m_Z$, the decay rate of the axino into a photon and a neutrino is given by
\begin{equation}
	\Gamma(\tilde a \to \gamma \nu_i) \simeq \frac{C_W^2\alpha_W^2}{128 \pi^3}\frac{m_{\tilde a}^3}{f_a^2}
	\left(\frac{g_2^2 \langle\tilde \nu_i\rangle^2}{2m_{\tilde W}^2}\sin^2\theta_W \right),
\end{equation}
where the factor $(g_2\langle\tilde \nu_i\rangle/\sqrt{2})/m_{\tilde W}$ in the parenthesis is the mixing between the wino and the neutrino, and $\sin\theta_W$ the mixing between $W^0$ and the photon.
Similarly, the decay rate of $\tilde a \to Z \nu_i$ becomes
\begin{equation}
	\Gamma(\tilde a \to Z \nu_i) \simeq \frac{C_W^2\alpha_W^2}{128 \pi^3}\frac{m_{\tilde a}^3}{f_a^2}
	\left(\frac{g_2^2 \langle\tilde \nu_i\rangle^2}{2 m_{\tilde W}^2}\cos^2\theta_W \right)
	\left(1- \frac{m_Z^2}{m_{\tilde a}^2} \right)
	\left( 1-\frac{m_Z^2}{2m_{\tilde a}^2} - \frac{m_Z^4}{2m_{\tilde a}^4} \right),
\end{equation}
while that of $\tilde a \to l_i W$, which is a sum of the rates of $\tilde a \to l_i^+ W^-$ and $\tilde a \to l_i^- W^+$, is
\begin{equation}
	\Gamma(\tilde a \to W l_i) \simeq \frac{C_W^2\alpha_W^2}{128\pi^3}\frac{m_{\tilde a}^3}{f_a^2}
	\left(\frac{g_2^2 \langle\tilde \nu_i\rangle^2}{m_{\tilde W}^2} \right)
	\left(1- \frac{m_W^2}{m_{\tilde a}^2} \right)
	\left( 1-\frac{m_W^2}{2m_{\tilde a}^2} - \frac{m_W^4}{2m_{\tilde a}^4} \right),
\end{equation}
where $m_W$ is the mass of the $W$ boson, and the factor $g_2\langle\tilde \nu_i\rangle/m_{\tilde W}$
represents the mixing between the charged wino and the lepton.
Thus, we obtain ${\rm Br} (\tilde a \rightarrow \gamma \nu) : {\rm Br} (\tilde a \rightarrow Z \nu) :
{\rm Br} (\tilde a \rightarrow Wl)
\simeq \sin^2 \theta_W : \cos^2 \theta_W : 2$ for $m_{\tilde a} \gg m_Z, m_W$. 
This results in the branching fraction of $\tilde a \to \gamma \nu$ of around $0.09$.

\begin{figure}
\begin{center}
\vskip -1.cm
\includegraphics[scale=1.2]{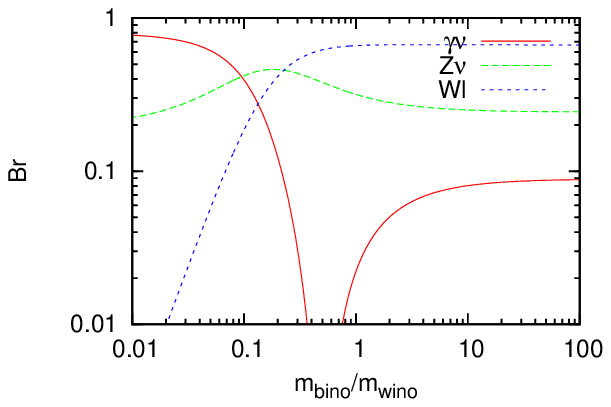}
\includegraphics[scale=1.2]{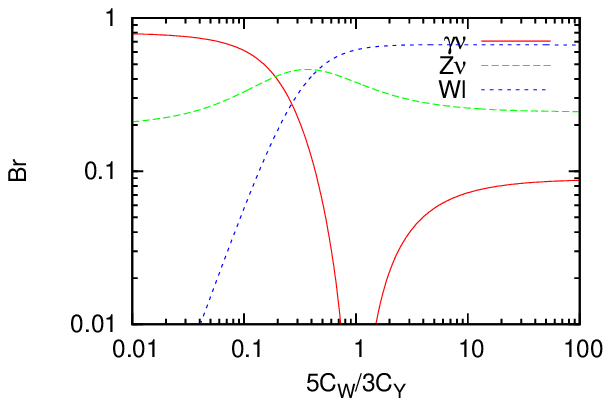}
\vskip -0.1cm
\caption{The branching ratios of $\tilde a$ into $\gamma \nu$  ({\color{red} red}), 
$Z \nu$  ({\color{green} green}) and $W l$  ({\color{blue} blue}) as a function of $m_{\tilde B}/m_{\tilde W}$
for $3C_Y/5 = C_W$ (left) and as a function of $5C_W/3C_Y$ for $m_{\tilde B}/m_{\tilde W} = 0.5$ (right).
We assume $m_{\tilde a} \simeq 260\,$GeV and $|m_{\tilde B} - \mu|, |m_{\tilde W} - \mu| \gg m_Z$.}
\label{fig:br_mass}
\end{center}
\end{figure}

In the case of $C_Y \sim C_W$, both the first and second terms in Eq.~\eqref{aYY} contribute
to the axino decay. The decay rates in such a generic case are summarized in App.~\ref{sec:decay_rate}.
As shown there, the branching fractions are determined by a combination of
$(m_{\tilde B}/m_{\tilde W})(C_W/C_Y)$ for $|m_{\tilde W} - \mu|, |m_{\tilde B} - \mu|\gg m_Z$.
Fig.~\ref{fig:br_mass} shows the branching ratios of the axino decay into $\gamma \nu$, $Z \nu$ and $W l$
as a function of $m_{\tilde B}/m_{\tilde W}$ for $C_W = 3C_Y/5$ (left)
and as a function of $5C_W / 3C_Y$ for $m_{\tilde B}/m_{\tilde W}=0.5$ (right).
One can confirm that the branching ratio of $\tilde a \to \gamma \nu$ becomes $0.8$
for $(m_{\tilde B}/m_{\tilde W})(C_W/C_Y)\to 0$,
while it becomes $0.09$ for large $(m_{\tilde B}/m_{\tilde W})(C_W/C_Y)$.
In the intermediate regime, the branching ratio of $\tilde a \to \gamma \nu$ decreases due to an
interference effect and eventually vanishes at $(m_{\tilde B}/m_{\tilde W})(C_W/C_Y)=\alpha_Y/\alpha_W$ 
[cf.~Eq.~(\ref{eq:decay_limit})]. In most of the parameter space, however, ${\rm Br} (\tilde a \rightarrow \gamma \nu) > O(0.01)$ and hence the model can explain the gamma-line without overproducing continuum gamma-ray and antiprotons.

\subsection{A model of SUSY axion}  \label{sec:model}

Here, we briefly describe an explicit model of the SUSY axion.
Let us introduce PQ superfields, $\Phi$ and $\bar\Phi$, with PQ charges of $+1$ and $-1$, respectively.
Also, PQ quarks, $Q$ and $\bar Q$, are added, which have fundamental and anti-fundamental representations of the SM SU(3), respectively, and both of which have a PQ charge of $-1/2$.
The superpotential is given by
\begin{equation}
	W_{\rm PQ} = \lambda X (\Phi\bar\Phi - V^2) + k \Phi Q\bar Q +W_0,
\end{equation}
where $X$ is a singlet superfield, $\lambda$ and $k$ are coupling constants, and $W_0 = m_{3/2} M_P^2$
is a constant term with the gravitino mass $m_{3/2}$ and the reduced Planck scale $M_P$.
The coupling constants are taken to be real and positive.
Including the SUSY breaking terms, $m_{\Phi}^2$ and $m_{\bar\Phi}^2$, 
the relevant terms of the scalar potential are
\begin{equation}
	V_{\rm PQ} = m_{\Phi}^2|\Phi|^2 + m_{\bar\Phi}^2|\bar\Phi|^2 
	+ \lambda^2 |\Phi\bar\Phi - V^2|^2 + \lambda^2 |X|^2( |\Phi|^2+ |\bar\Phi|^2 )
	+ (2 \lambda m_{3/2} V^2 X + {\rm h.c.}), \label{VPQ} 
\end{equation}
where we have assumed the minimal K\"ahler potential, for simplicity.
The VEVs of the PQ scalars are given by $\langle \Phi \rangle \sim \langle \bar\Phi \rangle \sim V$,
which is related to the PQ scale $f_a$ as $f_a = \sqrt{2 ( \langle \Phi \rangle^2+\langle \bar\Phi \rangle^2 )}$,
and the PQ quarks obtain a mass of $k V$.
The axion, which is a goldstone boson associated with the VEVs of the PQ scalars, has an anomaly-induced coupling
to the gluon via PQ quark loops, because the PQ symmetry is anomalous under the QCD.
Thus it solves the strong CP problem.
The coupling constants, $C_Y$ and $C_W$, depend on assignments of the U(1)$_Y$ and SU(2) charge on the PQ quarks.
If they are a singlet under SU(2) but have (opposite) U(1)$_Y$ charges, we obtain $C_W = 0$ and $C_Y \neq 0$.
If they are embedded in the SU(5) representation, both $C_Y$ and $C_W$ are nonzero and 
satisfy $C_W = 3C_Y/5$.
For instance, if the PQ quarks are embedded in ${\rm 5} + {\rm \bar 5}$ of SU(5),
the coefficients become $3C_Y/5 = C_W = 1$.

In this model, the axino, that is a fermionic component of a linear combination of $\Phi$ and $\bar\Phi$,
obtains a mass of $m_{\tilde a}= \lambda \langle X\rangle \simeq m_{3/2}$, where the VEV of $X$ is derived from \eqref{VPQ}. 
Several additional effects can make the axino heavier or lighter. 
There can be other SUSY breaking contributions to the tadpole term of $X$ in (\ref{VPQ}), 
which change the VEV of $X$ and hence the axino mass.
Radiative correction from the $Q (\bar Q)$ loops can also modify the axino mass~\cite{Goto:1991gq}.
In this letter, we assume that these effects slightly reduce the axino mass, and the axino becomes the LSP.

\subsection{Cosmology}  \label{sec:cos}

In this section, we discuss several cosmological constraints on the decaying axino DM scenario.

\subsubsection{Lightest neutralino}  

Let us assume that the lightest neutralino is mostly composed of the bino 
and that it is the MSSM-LSP.
In the presence of the R-parity violation, the bino decays into $Z\nu_i, W^\pm l_i^\mp$ and $h\nu_i$ due to the sneutrino VEV.
The decay rate of the bino is given by~\cite{Ishiwata:2008cu}
\begin{equation}
	\frac{1}{\Gamma_{\tilde B}^{\rm (RPV)}} \simeq 2\times 10^{-3}\,{\rm sec}\left( \frac{\kappa}{10^{-11}} \right)^{-2}
	\left( \frac{m_{\tilde B}}{1\,{\rm TeV}} \right)^{-1}.
\end{equation}
In order for the bino decay not to disturb the BBN, i.e., for the bino lifetime shorter than 0.1\,sec, we need $\kappa \gtrsim 10^{-12}$ for $m_{\tilde B} = 1\,{\rm TeV}$.

The bino also decays into the axino through the R-parity conserving operators (\ref{aYY}) as $\tilde B \to \tilde a \gamma$ and $\tilde B \to \tilde a Z$.
If this dominates the bino decays, the produced axinos may exceed the observed DM abundance.
In order to avoid the axino overproduction, the production rate
should be much less than $\Gamma_{\tilde B}^{\rm (RPV)}$.
The decay rate of the bino into axinos with photons or $Z$ bosons is totally given by~\cite{Covi:2001nw}
\begin{equation}
	\frac{1}{\Gamma_{\tilde B}^{\rm (PQ)}} \simeq 3\times 10^{2}\,{\rm sec}~C_Y^{-2}
	\left( \frac{f_a}{10^{15}\,{\rm GeV}} \right)^{2}
	\left( \frac{m_{\tilde B}}{1\,{\rm TeV}} \right)^{-3}.
\end{equation}
The axino abundance produced by the bino decay in terms of the density parameter 
$\Omega_{\tilde a} \equiv \rho_{\tilde a}/\rho_{\rm cr}$, where $\rho_{\tilde a}$ is the present energy density of the axino and $\rho_{\rm cr}$ is the present critical energy density, becomes
\begin{equation}
	\Omega_{\tilde a}^{(\tilde B)} h^2 =  \frac{m_{\tilde a}}{m_{\tilde B}} 
	\frac{\Gamma_{\tilde B}^{\rm (PQ)}}{\Gamma_{\tilde B}^{\rm (PQ)} + \Gamma_{\tilde B}^{\rm (RPV)}}
	\Omega_{\tilde B} h^2,
	\label{Omega_a_B}
\end{equation}
where $h$ is the present Hubble parameter in units of 100\,km/s/Mpc,
and $\Omega_{\tilde B}$ is the bino abundance after the thermal decoupling evaluated as if the bino were stable.
Since the bino abundance is large in general, the axion abundance becomes too large,
unless the branching ratio of the bino decay into the axino is suppressed, i.e., $\Gamma_{\tilde B}^{\rm (RPV)} \gg \Gamma_{\tilde B}^{\rm (PQ)}$.

\subsubsection{Axino and axion}  

Axinos are produced by scatterings of the gluons and the gluinos from the thermal bath at the reheating.
The thermally produced axino abundance is given by~\cite{Brandenburg:2004du,Strumia:2010aa}
\begin{equation}
	\Omega_{\tilde a}^{\rm (th)} h^2 \simeq 6\times 10^{-3} g_3^6 \ln\left( \frac{3}{g_3} \right)
	\left( \frac{m_{\tilde a}}{260\,{\rm GeV}} \right)
	\left( \frac{T_{\rm R}}{10^6\,{\rm GeV}} \right)
	\left( \frac{f_a}{10^{15}\,{\rm GeV}} \right)^{-2},
\end{equation}
where $g_3$ is the running QCD coupling constant at the $T_{\rm R}$ scale, and $T_{\rm R}$ is the reheating temperature after the inflation. This is valid as long as $T_{\rm R}$ is larger than the axino mass. 
Thus, the axino can be a dominant component of the DM for
$T_{\rm R}\sim 10^{5}\,{\rm GeV}(f_a/10^{14}\,{\rm GeV})^2$.

The abundance of the axion coherent oscillation is estimated as~\cite{Turner:1985si}
\begin{equation}
	\Omega_a h^2 \simeq 0.2 \theta_a^2 \left( \frac{f_a}{10^{12}\,{\rm GeV}} \right)^{1.19},
\end{equation}
where $\theta_a$ denotes the axion initial misalignment angle.
For $f_a\sim 10^{13}-10^{14}$\,GeV, we need $\theta_a \lesssim O(0.1)$ in order for the axion abundance to be lower than the DM abundance.

\subsubsection{Saxion}  

The saxion, $\sigma$, belongs to a flat direction in the scalar potential (\ref{VPQ}), which satisfies $\Phi\bar\Phi = V^2$.
It obtains a mass, $m_\sigma$, from the SUSY breaking effect.

Let us estimate the saxion abundance. 
The saxion sits around the minimum during the inflation, which is slightly displaced from the low-energy true minimum,
and begins to oscillate around the true minimum when the Hubble parameter decreases to $m_\sigma$ with an initial amplitude of $\sigma_i$. The abundance of saxion coherent oscillation is
\begin{equation}
\begin{split}
	\frac{\rho_\sigma}{s} = \frac{1}{8}T_{\rm R}\left( \frac{\sigma_i}{M_P} \right)^2
	\simeq 2\times 10^{-2}\,{\rm GeV}
	\left( \frac{T_{\rm R}}{10^6\,{\rm GeV}} \right)
	\left( \frac{f_a}{10^{15}\,{\rm GeV}} \right)^{2}
	\left( \frac{\sigma_i}{f_a} \right)^2,
\end{split}
\end{equation}
where $\rho_\sigma$ is the saxion energy density, and $s$ is the entropy density.
Here, we have assumed that the saxion oscillation starts before the reheating process of the inflation is finished, which is the case for $T_{\rm R}\lesssim 10^{10}\,{\rm GeV} (m_\sigma/500\,{\rm GeV})^{1/2}$.
The saxion dominantly decays into the axion pair. The lifetime becomes~\cite{Chun:1995hc}
\begin{equation}
	\tau_\sigma =  \left( \frac{\xi^2}{32\pi}\frac{m_\sigma^3}{f_a^2} \right)^{-1} \simeq 0.5\,{\rm sec}
	\left( \frac{f_a}{10^{15}\,{\rm GeV}} \right)^{2}
	\left( \frac{m_{\sigma}}{500\,{\rm GeV}} \right)^{-3} \xi^{-2},
\end{equation}
with $\xi \equiv 2(\langle\Phi\rangle^2-\langle\bar\Phi\rangle^2)/f_a^2$.
Note that the saxion also decays into a pair of the gluons with a branching fraction of $\sim (g_3^2/4\pi^2)^2$~\cite{Ichikawa:2007jv}, but it 
does not affect the BBN as long as $f_a \lesssim 10^{15}$\,GeV is satisfied for $m_\sigma \sim 500$\,GeV.\footnote{
	We assume that the saxion is lighter than twice the mass of the axino. 
	Otherwise, the axino LSPs are overproduced by the saxion decay, $\sigma \to \tilde a \tilde a$. This indicates $m_\sigma < 520$\,GeV.
}

The axions produced by the saxion decay contribute to the extra effective number of the neutrino species, $\Delta N_{\rm eff}$
\cite{Choi:1996vz,Chun:2000jr,Ichikawa:2007jv,Kawasaki:2011ym,Kawasaki:2011rc,Jeong:2012np,Moroi:2012vu,Choi:2012zna},
which is given by
\begin{equation}
	\Delta N_{\rm eff} = \frac{43}{7}\left[\frac{10.75}{g_{*s}(T_\sigma)}\right]^{1/3} 
	\left(\frac{\rho_\sigma}{\rho_{\rm rad}}\right)_{T=T_\sigma},
\end{equation}
where 
\begin{equation}
	\left(\frac{\rho_\sigma}{\rho_{\rm rad}}\right)_{T=T_\sigma} = \frac{T_{\rm R}}{6T_{\rm \sigma}}\left( \frac{\sigma_i}{M_P} \right)^2,
\end{equation}
with $T_\sigma$ denoting the temperature at which the saxion decays and $\rho_{\rm rad}$ being the radiation energy density.
It is estimated as
\begin{equation}
	\Delta N_{\rm eff} \simeq 1.2
	\left( \frac{f_a}{10^{15}\,{\rm GeV}} \right)^3
	\left( \frac{T_{\rm R}}{10^{4}\,{\rm GeV}} \right)
	\left( \frac{m_\sigma}{500\,{\rm GeV}} \right)^{-3/2}
	\left( \frac{\sigma_i}{f_a} \right)^2 \xi^{-1}.
	\label{Neff}
\end{equation}
The contribution should satisfy $\Delta N_{\rm eff} \lesssim 1$.
In other words, the recent claims of the existence of the extra light species,
$\Delta N_{\rm eff}\simeq 1$~\cite{Hou:2012xq}, can be explained by the non-thermal axions from the saxion decay.
We assume $\xi\sim1$ for simplicity.

\begin{figure}
\begin{center}
\includegraphics[scale=1.2]{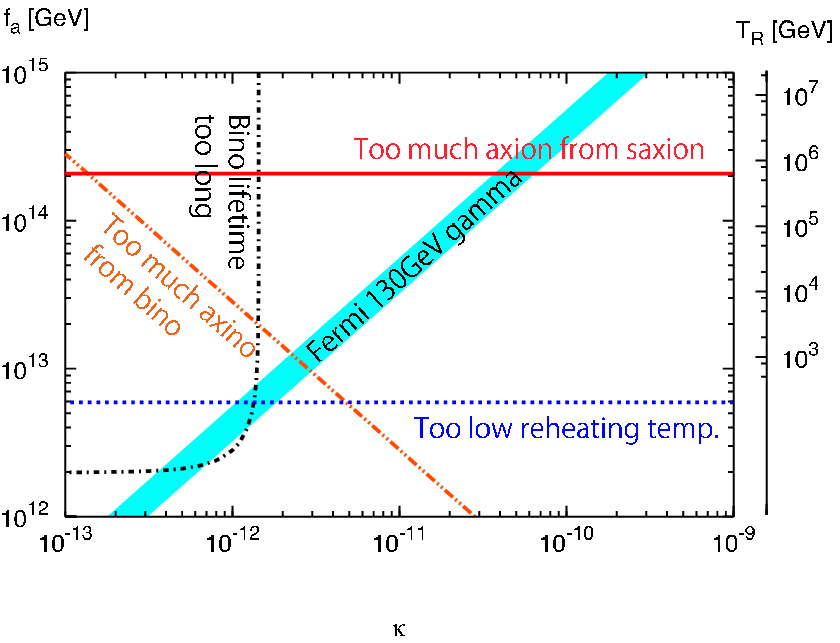}
\includegraphics[scale=1.2]{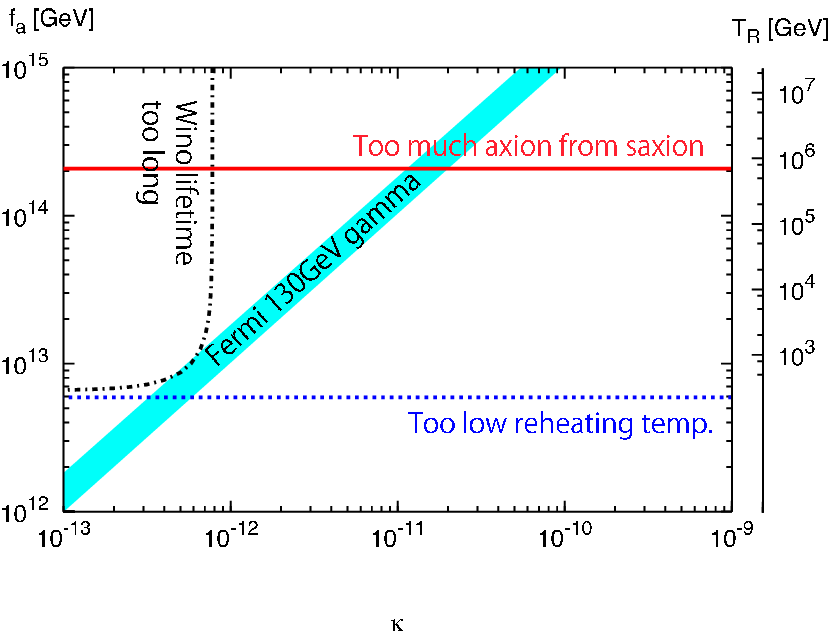}
\caption{(Top) Constraints on the plane of $\kappa$ and $f_a$ 
for the bino MSSM-LSP with $C_W=0$, $C_Y=1$,
$m_{\tilde a}=260$\,GeV,
$m_{\tilde B}=1$\,TeV, $m_\sigma=500$\,GeV and $\sigma_i = f_a$.
In the light blue band, the Fermi 130\,GeV gamma-ray line excess is explained.
On the right side of the black dashed line, the bino lifetime is shorter than $0.1$\,sec.
Above the orange dot-dashed line, the axino abundance produced by the bino decay satisfies 
$\Omega_{\tilde a}^{(\tilde B)} h^2 < 0.1$, where $\Omega_{\tilde B}h^2 = 10$ is taken as a reference.
For each $f_a$, $T_{\rm R}$ is set so that the axino becomes the dominant component of the DM.
Below the red horizontal line, $\Delta N_{\rm eff} < 1$ is obtained from the axions produced by the saxon decay.
Above the blue dotted horizontal line, $T_{\rm R} > m_{\tilde a}$ is satisfied to account for the axino DM.
(Bottom) Same as the top panel, but for the wino MSSM-LSP with 
$C_W=1$, $C_Y=0$ and  $m_{\tilde W}=1$\,TeV.}
\label{fig:cont}
\end{center}
\end{figure}

\subsubsection{Combined constraints}

The combined constraints on a plane of $(\kappa, f_a)$ are shown in Fig.~\ref{fig:cont}.
In the top panel, we have taken $C_W=0$, $C_Y=1$,
$m_{\tilde a}=260$\,GeV,
$m_{\tilde B}=1$\,TeV, $m_\sigma=500$\,GeV and $\sigma_i = f_a$.
In the light blue band, the Fermi 130\,GeV gamma-ray line excess is explained.
On the right side of the black dashed line, the bino lifetime is shorter than $0.1$\,sec, and the decay has no effects on the BBN.
Above the orange dot-dashed line, we have $\Omega_{\tilde a}^{(\tilde B)} h^2 < 0.1$,
and the axino abundance produced by the bino decay is sufficiently small.
Here, $\Omega_{\tilde B}h^2 = 10$ is taken as a reference value of the bino abundance as inferred from Ref.~\cite{Kawasaki:2008qe}.
Below the red horizontal line, $\Delta N_{\rm eff} < 1$ is obtained from the axions produced by the saxion decay.
In the figure, $T_{\rm R}$ is set so that $\Omega_{\tilde a}^{(\rm th)} h^2 = 0.1$ is realized for each $f_a$. 
Above the blue dotted horizontal line, $T_{\rm R} > m_{\tilde a}$
and the axino DM is thermally produced.

In the bottom panel, we have taken $C_W=1$ and $C_Y=0$ and assumed that the wino is the MSSM-LSP
with its mass of $m_{\tilde W}=1$\,TeV. 
Note that since the thermal relic wino abundance is small, there is no constraint from the axino overproduction by the wino decay.\footnote{Similar conclusions hold for the case of the stau MSSM-LSP.}
In both cases, it is found that the Fermi 130\,GeV gamma-ray line excess is accounted for without suffering from the cosmological constraints for 
$\kappa \simeq O(10^{-12}-10^{-11})$ and $f_a \simeq 10^{13}-10^{14}$\,GeV.

Here we briefly mention parameter dependences on these constraints.
For larger neutralino mass, the axino lifetime becomes longer and the light-blue band moves to the bottom-right.
Also, the axino abundance from the neutralino decay (\ref{Omega_a_B}) becomes larger 
and the orange dot-dashed line moves to the top-right.
For smaller saxion mass or larger initial amplitude of the saxion,
the bound from $N_{\rm eff}$ (\ref{Neff}) becomes stronger and the red line moves to the bottom.
For general values of $C_W$ and $C_Y$, the branching ratio of the axino decay into photon
may become smaller, as shown in Fig.~\ref{fig:br_mass}, and hence the light-blue band moves to the bottom-right
to make the axino lifetime smaller.

Before closing this section, let us comment on the gravitino.
The gravitinos are also produced by scatterings of the gluons and the gluinos from the thermal bath at the reheating.
If the gravitino is lighter than the MSSM-LSP,
then it dominantly decays into the axino and the axion, 
and hence there is no BBN constraint~\cite{Asaka:2000ew}.
The nonthermal production of axinos by the gravitino decay is negligible,
since the abundance of the thermally produced gravitino is less than that of the axino.
On the other hand, if the gravitino is heavier than the MSSM-LSP, its decay affects the BBN.
The parameter range corresponding to $T_{\rm R}\simeq O(10^5)$\,GeV (just below the red horizontal line in Fig.~\ref{fig:cont}) is constrained depending on the gravitino mass and the MSSM mass spectrum~\cite{Kawasaki:2008qe}.

\section{Conclusion} 
\label{sec:conc}

We have proposed the decaying axino DM scenario 
as a model to explain the Fermi 130\,GeV gamma-ray line excess from the GC.
It is based on the SUSY KSVZ axion model with the bilinear R-parity violation.
The model realizes a fairly large branching fraction of the axino decay into a photon plus a neutrino. 
It was found that the Fermi excess is accounted for while satisfying the other cosmological constraints 
for $f_a \simeq 10^{13}-10^{14}$\,GeV and $\kappa \simeq O(10^{-12}-10^{-11})$. 

Compared to another well--motivated decaying DM, i.e., the decaying gravitino DM,
the decaying axino DM typically has a larger branching fraction into the monochromatic gamma.
The gravitino universally couples to the lepton and the Higgs superfields,
and hence the gravitino's decay into $h \nu$, $Z\nu$ and $W l$ cannot be suppressed
in the presence of the bilinear R-parity violation.
Thus, the decaying gravitino DM is severely constrained by the observation of the antiproton flux~\cite{Buchmuller:2012rc}.

Let us touch on the collider phenomenology. The MSSM-LSP is stable in the detectors for the R-parity violation of $\kappa \sim 10^{-11}$ (see e.g., Ref.~\cite{Asai:2011wy}). Thus, when the MSSM-LSP is neutral, the SUSY events would be detected by searching for signals with a large missing transverse momentum. If the MSSM-LSP is a charged particle, it leaves a charged track, which is a characteristic signal in the detector. The sensitivities of the LHC searches are the same as the standard SUSY searches. 

As mentioned in the introduction, the morphology of the observed gamma-ray signature is still premature to refute the decaying DM scenario. If the uncertainties will be understood in future, such an analysis can be used for distinguishing the DM models particularly between the decaying and annihilating DM models. 
If the former scenario will become favored, the decaying axino DM model can be an attractive candidate.

\section*{Acknowledgment}

This work was supported by JSPS KAKENHI Grant 
No.~23740172 (M.E.), No.~21740164 (K.H.), No.~22244021 (K.H.),
No.\ 22244030 (K.N.) and by the MEXT Grant-in-Aid No.\ 21111006 (K.N.).
The work of K.M. is supported in part by JSPS Research Fellowships
for Young Scientists.
This work was supported by World Premier International Research Center Initiative (WPI Initiative), MEXT, Japan.

\appendix

\section{Formulae for the axino decay rate}
\label{sec:decay_rate}

In this appendix, we provide the general formulae for the axino decay rate in the presence of the R-parity violation.
The interaction between the axion and the SU(2)$\times$U(1) gauge supermultiplets is given by
\begin{equation}
\mathcal{L} = \frac{\alpha_Y C_Y}{4 \sqrt{2} \pi f_a} \int d^2\theta A \,\mathcal{W}_{B} \mathcal{W}_{B} + 
\frac{\alpha_W C_W}{4 \sqrt{2} \pi f_a} \int d^2\theta A \,\mathcal{W}^{a}_{W} \mathcal{W}^{a}_{W} + {\rm h.c.},
\end{equation}
where $\alpha_{Y(W)}$ stands for the fine structure constant of U(1)$_{Y}$ (SU(2)), 
$f_a$ is the PQ scale, $A$ is the axion superfield, $A = (\sigma + ia)/\sqrt{2} + \sqrt{2}\tilde a \theta + F^A\theta^2$, and $\mathcal{W}_{B}$ ($\mathcal{W}^{a}_{W}$) is the supersymmetric field strength of U(1) (SU(2)). 
The coefficients $C_Y$ and $C_W$ are model-dependent constants of order unity.
In terms of the component fields, 
the axino interactions become
\begin{align}
	 \mathcal{L} =  i \frac{\alpha_Y C_Y}{16 \pi f_a}\bar{\tilde{a}}
	 \gamma_5[\gamma^{\mu},\gamma^{\nu}]\tilde{B}B_{\mu \nu} + i \frac{\alpha_W C_W}{16 \pi f_a}
	 \bar{\tilde{a}}\gamma_5[\gamma^{\mu},\gamma^{\nu}]\tilde{W}^aW^a_{\mu \nu},
	 \label{eq:int}
\end{align}
where $\tilde a$ is the axino, $B_{\mu \nu}$ ($W_{\mu \nu}$) is the gauge boson of U(1)$_Y$ (SU(2)), and $\tilde B$ and $\tilde W^a$ denote the bino and the wino, respectively.

Due to the sneutrino VEV induced by the bilinear R-parity violation, 
the SM leptons and the gauginos mix with each other,
and the axino LSP, $\tilde a$, can decay into $\gamma \nu$, $Z \nu$ and $W l$.
The mass matrix of the neutralino and the neutrino becomes
\begin{align}
	 M_N=
 	\left(\begin{array}{ccccc} 
 	m_{\tilde B} & 0 & -m_Zs_Wc_\beta & m_Zs_Ws_\beta & - g_Y \<\tilde \nu_i\> / \sqrt{2}\\ 
 	0 & m_{\tilde W} & m_Zc_Wc_\beta &-m_Zc_Ws_\beta & g_2 \< \tilde \nu_i \>/ \sqrt{2}\\
 	-m_Zs_Wc_\beta & m_Zc_Wc_\beta & 0 & -\mu & 0 \\
 	m_Zs_Ws_\beta & -m_Zc_Ws_\beta & -\mu & 0 &0 \\
	- g_Y \<\tilde \nu_i\> / \sqrt{2} & g_2 \< \tilde \nu_i \>/ \sqrt{2} & 0 & 0 & 0\\
	\end{array} \right),
	\label{eq:mass_N}
\end{align}
for $\mathcal L_m = -\frac{1}{2} (\tilde\psi^0)^T M_N \tilde\psi^0 + {\rm h.c.}$ 
with $\tilde\psi^0 = (\tilde B, \tilde W^0, \tilde H_d^0, \tilde H_u^0, \nu_i)^T$. 
Here, the subscript $i = 1,2,3$ stands for the generation, 
$c_\beta=\cos\beta$, $s_\beta=\sin\beta$, 
$c_W=\cos\theta_W$
and
$s_W=\sin\theta_W$
with the weak mixing angle $\theta_W$.
The upper-left $4\times 4$ matrix is 
identical to the neutralino mass matrix of the MSSM, and the neutrino masses are approximated to be zero.
On the other hand, the mass matrix of the chargino and the lepton is given by
\begin{align}
	M_C = \left( 
	\begin{array}{ccc} 
	m_{\tilde W} & \sqrt{2}m_Ws_{\beta} &0 \\
	\sqrt{2}m_Wc_{\beta} & \mu & - Y^l_{i} \< \tilde \nu_i \> \\
	g_2\< \tilde \nu_i \> & 0 & m_{l_i}
	\end{array} \right),
	\label{eq:mass_C}	
\end{align}
for $\mathcal L_m = -(\tilde\psi^-)^T M_C \tilde\psi^+ + {\rm h.c.}$ 
with $\tilde\psi^- = (\tilde W^-, \tilde H_d^-, l_{Li}^-)^T$ 
and $\tilde\psi^+ = (\tilde W^+, \tilde H_u^+, l_{Ri}^+)^T$. 
Here, $Y^l_{i}$ is the lepton Yukawa coupling constant, and 
the upper-left $2\times 2$ matrix is identical to the chargino mass matrix of the MSSM.
Neglecting the small SM lepton masses
in the phase space, one finds the decay rates as
\begin{align}
	\Gamma (\tilde a \rightarrow GL) = \frac{1}{128 \pi^3}
	\frac{m_{\tilde{a}}^3}{f_a^2} \left(1-\frac{m_G^2}{m_{\tilde a}^2}\right) 
	\left(1 - \frac{m_G^2}{2 m_{\tilde a}^2 }
	- \frac{m_G^4}{2 m_{\tilde a}^4} \right)
	F_{GL} (C_Y, C_W, m_{\tilde B}, m_{\tilde W})
\end{align}
with
\begin{align}
	F_{GL}& (C_Y, C_W, m_{\tilde B}, m_{\tilde W}) \equiv \nn
	&
	\begin{cases}
		\left|
		\alpha_Y C_Y \cos \theta_W U_{\nu_i \tilde B} +
		\alpha_W C_W \sin \theta_W U_{\nu_i \tilde W}
		\right|^2&\mbox{for}~(G,L)=(\gamma, \nu_i), \\[5pt]
		\left| -  \alpha_Y C_Y \sin \theta_W U_{ \nu_i \tilde B}
		+ \alpha_W C_W \cos \theta_W U_{ \nu_i \tilde W} \right|^2&\mbox{for}~(G,L)=(Z, \nu_i), \\[5pt]
	\alpha_W^2 C_W^2 \( 
	\left| U_{l_i \tilde W}^L  \right|^2 
	+ \left| U_{l_i \tilde W}^R \right|^2  \).
	&\mbox{for}~(G,L) = (W, l_i).
	\end{cases}
	\label{eq:decay_each}
\end{align}
where
$m_G \,(G=\gamma, Z, W)$ is a mass of the gauge boson, 
and $U_{\nu_i \tilde B}$, $U_{\nu_i \tilde W}$ and $U_{l_i \tilde W}^{L/R}$
are the mixings that diagonalize the mass matrices of Eqs.~(\ref{eq:mass_N}) and (\ref{eq:mass_C}).

In the limit of $|m_{\tilde B} - \mu|, |m_{\tilde W} - \mu| \gg m_Z$, 
the mixings between the gauginos and the higgsinos become irrelevant. 
Assuming that the R-parity violation is small, $U_{\nu_i \tilde B}$, $U_{\nu_i \tilde W}$ and $U_{l_i \tilde W}^{L/R}$ in Eq.~\eqref{eq:decay_each} 
are approximated as
\begin{align}
	F_{GL}& (C_Y,C_W,m_{\tilde B},m_{\tilde W})\simeq \nn
	&\begin{cases}
		\left|
		\alpha_Y C_Y \cos \theta_W \cfrac{g_Y \< \tilde \nu_i \>}{\sqrt{2} m_{\tilde B}} -
		\alpha_W C_W \sin \theta_W \cfrac{g_2 \< \tilde \nu_i \>}{\sqrt{2} m_{\tilde W}}
		\right|^2	
		&\mbox{for}~(G,L) = (\gamma,\nu_i), \\[20pt]
		\left| \alpha_Y C_Y \sin \theta_W \cfrac{g_Y \< \tilde \nu_i \>}{\sqrt{2} m_{\tilde B}}
		+ \alpha_W C_W \cos \theta_W \cfrac{g_2 \< \tilde \nu_i \>}{\sqrt{2} m_{\tilde W}} \right|^2		
		&\mbox{for}~(G,L) = (Z,\nu_i), \\[20pt]
		\alpha_W^2 C_W^2 \left| \cfrac{g_2 \< \tilde \nu_i \>}{m_{\tilde W}} \right|^2
		&\mbox{for}~(G,L) = (W,l_i),
	\end{cases}
	\label{eq:decay_limit}
\end{align}
where $\<\tilde \nu_i\>$ is the sneutrino VEV induced by the R-parity violation. 

As can be seen from \eqref{eq:decay_limit},
the branching fractions are determined by $(m_{\tilde B}/m_{\tilde W})(C_W/C_Y)$.
When either $C_W$ or $C_Y$ is sufficiently small, the decay rates are reduced to the expressions in Sec.~\ref{sec:R}.
If the PQ quarks are embedded in a complete vector-like multiplet of the GUT,
they become $C_W = 3C_Y/5$.
Moreover, if the GUT relation is assumed for the gaugino masses, they satisfy $m_{\tilde W}/m_{\tilde B} = 
3\alpha_W/5\alpha_Y$ at the one-loop level of the renormalization group evolution. 
Then, the decay rate of the axino to $\gamma \nu$ is found to vanish.
This cancellation is seen in Fig.~\ref{fig:br_mass}.



\end{document}